\documentclass[11pt]{article}

\usepackage[utf8]{inputenc}
\usepackage{graphicx}
\usepackage{authblk}	
\usepackage{amsmath,amsthm,amsfonts,amssymb,amscd}
\usepackage[margin=1.0in]{geometry}
\usepackage{hyperref}
\usepackage{natbib}
\setcitestyle{numbers,square}

\title{\bfseries On the thermodynamics-based equilibrium beach profile derived by Jenkins and Inman}

\author[1]{Sergio Maldonado\footnote{Correspondance: s.maldonado@soton.ac.uk}}
\author[2]{Marylin Uchasara\footnote{Currently at Laboratoire de Dynamique Hydro-S\'{e}dimentaire, Ifremer, Plouzan\'{e}, France}}
\affil[1]{\footnotesize Faculty of Engineering and Physical Sciences, University of Southampton, Southampton SO16 7QF, U.K.}
\affil[2]{Department of Physics, Paris-Sud University, 91400 Orsay Cedex, France}

\date{}

\begin{document}

\maketitle

\begin{abstract}
Based on the second law of thermodynamics, Jenkins and Inman (2006 \textit{J. Geophys. Res.}, \textbf{111}, C02003) claimed that an equilibrium beach profile described by an elliptic cycloid maximises the rate of wave energy dissipation. 
However, here we i) highlight that the solution proposed by Jenkins and Inman (the elliptic cycloid) is difficult to recover due to important information being absent; and ii) show that, in fact, other curves can be proposed (e.g. a line) that yield larger rates of energy dissipation as formulated by the aforementioned authors, thus invalidating their claim. 
Combined, these two crucial aspects associated with the reproducibility and validity of the research invite further scrutiny of the work and conclusions reached by Jenkins and Inman (2006). 
This paper also serves as an appendix to Maldonado (2020 \textit{J. Geophys. Res.-Oceans} \textbf{125}, e2019JC015876. \href{https://doi.org/10.1029/2019JC015876}{doi: 10.1029/2019JC015876}).  
\end{abstract}

\section{Context}
Energy-dissipation-based assumptions have been employed in the past to derive analytically curves describing equilibrium beach profiles (EBPs) (see e.g. \cite{Dean,Larson}). 
However, to the best of the authors' knowledge, Jenkins and Inman \cite{Jenkins} were the first to hypothesise that EBPs may adopt shapes that \textit{maximise} the rate of energy dissipation of both breaking and non-breaking waves. Interestingly, this hypothesis diametrically opposes that by \cite{Larson} for the latter case. Our focus here is on profiles under non-breaking waves because of their relevance in Maldonado \cite{Maldonado}, for which this paper serves as an appendix, and where it is argued that EBPs under non-breaking waves tend ot minimise (not maximise) wave energy dissipation. Jenkins and Inman  \cite{Jenkins} arrive at their hypothesis via the maximum entropy production formulation of the second law of thermodynamics, supplemented by certain assumptions pertaining to the shorezone system (e.g. that the system is isothermal). 
By means of linear wave theory, Jenkins and Inman \cite{Jenkins} formulate an integral associated with the dissipation of wave energy in profiles under non-breaking waves (therein referred to as `shorerise profiles'), for which a maximum is sought; namely (eq. 19 in their paper):

\begin{equation} \label{eq:integral}
 \int h^{-3 (n + 1)/4} \sqrt{1 + (x')^2} dh , 
\end{equation}

%
where $h$ is the local water depth, which varies with cross-shore distance, $x$ (note that $h=h(x)$ deifnes the equilibrium beach profile); $x' \equiv dx/dh$ is the reciprocal of the local bed slope, $dh/dx$; and $n$ is some positive constant that characterises the variation of the bed shear stress magnitude, $\tau_o$, with the flow velocity amplitude at the bed, $u_m$, according to $\tau_o \propto u_m^n$.

Thus, Jenkins and Inman reduce the problem to that of finding an equilibrium beach profile, given by the function $h(x)$, that maximises the above integral, where the limits of integration are the boundaries of the shorerise profile. 
Then, Jenkins and Inman proceed to find a solution to the problem, using calculus of variations, and claim that an elliptic cycloid (i.e. the curve traced by the trajectory of a point on the perimeter of a rolling ellipse) represents the function $h(x)$ that maximises the integral \eqref{eq:integral}. 
Therefore, the hypothesis of Jenkins and Inman that we aim to scrutinise here is the following: an equilibrium beach profile described by an elliptic cycloid maximises the rate of energy dissipation of non-breaking waves, in turn associated with the integral (\ref{eq:integral}).

\section{Critique}

Despite its novelty and promising results (calibrated elliptic cycloids do indeed agree well with the measured profiles considered), the work by Jenkins and Inman \cite{Jenkins} invites scrutiny and revision of several aspects, from the assumptions adopted to the mathematical derivations. However, we focus here on two specific points that, in our view, refute the hypothesis by Jenkins and Inman discussed above; namely:
\begin{enumerate}
\item Their proposed solution cannot be verified
\item Their proposed solution does not maximise the integral (\ref{eq:integral})
\end{enumerate}

The first point relates to the reproducibility of the research under consideration, while the second point is concerned with its validity.

\subsection{On the reproducibility of the research}

An objective of Jenkins and Inman \cite{Jenkins} is to find a function $x(h)$ --the inverse of $h(x)$-- that maximises the integral (\ref{eq:integral}). 
Therefore, the functional to be maximised is (the limits of integration are discussed in \S \ref{s:validity}):
\begin{equation} \label{eq:functional}
J[x(h)] = \int_{h_1}^{h_2} L[h;x,x'] dh  = \int_{h_1}^{h_2} h^{-3 (n + 1)/4} \sqrt{1 + (x')^2} dh .
\end{equation}

For eq. (\ref{eq:functional}) to attain a stationary value at $x(h)$, presumed by Jenkins and Inman to be a maximum, the Euler-Lagrange equation must be satisfied. 
This eventually reduces (see \cite{Jenkins} and \cite{Maldonado}) to solving the following integral:
\begin{equation}
\int \sqrt{\frac{\Omega h^\alpha }{1 - \Omega h^\alpha}} dh 
\end{equation}
(which is the dimensional version of eq. 21 in \cite{Jenkins}), where $\Omega$ is an integration constant and $\alpha = 3(n+1)/2$. 
Jenkins and Inman then ``rationalize the integrand (...) using two separate Euler substitutions (...)'', but, crucially, do not mention what these substitutions are. 
Moreover, the general solution provided, which has two roots, takes the following form (see eq. 22 in \cite{Jenkins}):
\begin{equation}
x = \frac{\Omega^{(\alpha - 1) / \alpha}}{\epsilon \sqrt{R}} \left[ - \sqrt{\frac{h^\alpha}{\Omega} - h^{2 \alpha}} + \frac{1}{2 \Omega} \arccos \left( 1 - 2 \Omega h^\alpha \right)   \right] ,
\end{equation}
where $\epsilon$ is, for our purposes, a constant. 
The first root of the solution is  then given as (see eq. 22a in \cite{Jenkins}):
\begin{equation}
R = \left( \frac{\pi}{2 I_e^{(2)}} \right)^2 \left[ 4 \Omega h^\alpha - 4 \Omega^2 h^{2 \alpha} + \frac{2}{1+\alpha} \left( 1 - 4 \Omega h^\alpha + 4 \Omega^2 h^{2 \alpha} \right)   \right] ,
\end{equation}
with the second root being similar in form but dependent on $I_e^{(1)}$, where ``$I_e^{(1)}$ and $I_e^{(2)}$ are elliptic integrals of the first and second kind, respectively''. 
However, Jenkins and Inman do not mention whether they refer to incomplete or complete elliptic integrals and, more importantly, do not give the argument(s) of these functions (see Appendix A), which precludes us from recovering their solution to the variational problem or verifying, analytically, that it is correct. 

The authors of this paper do not wish to conclusively assert that the above solution to the variational problem put forward by Jenkins and Inman is incorrect. But we do wish to highlight that, given that said solution is arguably a central contribution of Jenkins and Inman \cite{Jenkins}, the fact that arriving at it is made difficult (we have been unable to recover it) by omitting the details discussed above invites further scrutiny and fails to promote reproducibility of their derivations. On a very different basis (exploiting mathematical analogies between beach profiles and relativistic cosmology), Faraoni \cite{Faraoni} has also recently questioned the reproducibility of the solution by Jenkins and Inman. Faraoini \cite{Faraoni} points out that existing analytical solutions of the variational problem posed by \cite{Jenkins} can be found only for physically meaningless (in the context of beach profiles) values of $n$, such as $n<0$.

\subsection{On the validity of the solution} \label{s:validity}

Jenkins and Inman \cite{Jenkins} rewrite their solution (eq. 22 in their paper) in the form of  a curve describing an elliptic cycloid. The main calibration parameter is then the eccentricity of the ellipse, $e$, in turn related to $n$ in $\tau_o \propto u_m^n$ via (see eq. 28 in \cite{Jenkins}):
\begin{equation} \label{eq:e-n}
    e = \left[ 1 - \frac{4}{(3n + 5)} \right ]^{1/2} .
\end{equation}

To test whether the solution by Jenkins and Inman maximises \eqref{eq:functional}, we simply compare the value of the functional (\ref{eq:functional}) yielded by Jenkins and Inman's solution against that obtained from some arbitrarily neighbouring curves complying with the same boundary conditions. 
We do so for the seaward or shorerise part of the profile solely, as discussed previously. 
The arbitrary curves to be tested are as follows.

Curve A -- a linear profile. Reason for selection: simplicity (but see also section 5 in \cite{Maldonado}),
\begin{equation}
x(h) = a h + b .
\end{equation} 

Curve B -- from \cite{Maldonado}. Reason for selection: to test another curve that also depends on $n$,
\begin{equation}
x(h) = a h^{(3 n + 7)/4} + b .
\end{equation}

Curve C -- particular case of Curve B when $n=2$. Reason for selection: to test some arbitrary non-linear profile,
\begin{equation}
x(h) = a h^{13/4} + b .
\end{equation}

Values of the constants $a$ and $b$ in the above expressions are given by the boundary conditions $x(h=h_1) = x_1$ and $x(h=h_2) = x_2$, in turn obtained from Jenkins and Inman's solutions, as illustrated in fig. \ref{fig:profile} below. 

\begin{figure}[h]
	\centering
	\includegraphics[width=0.9\linewidth]{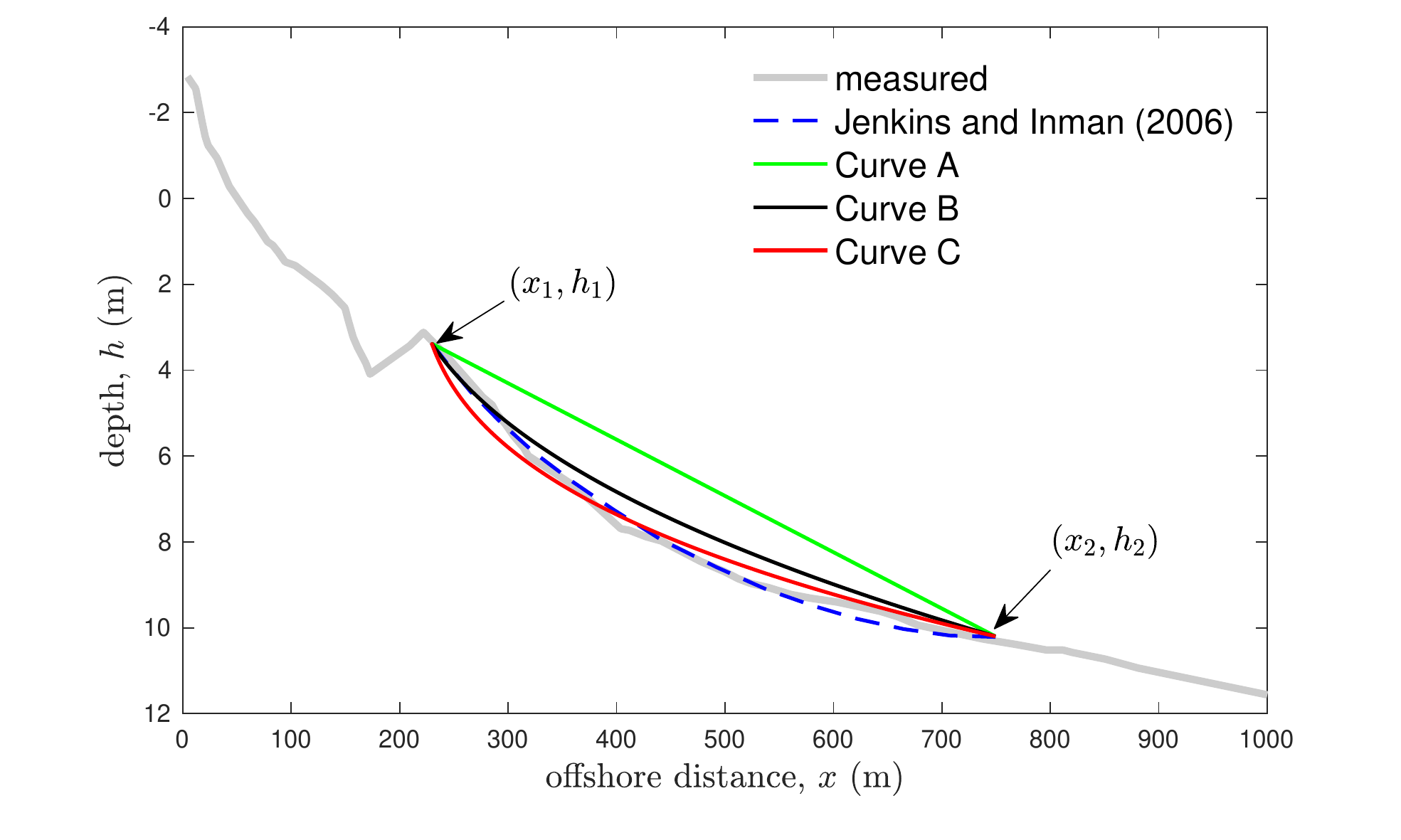}
	\caption{Comparison of Jenkins and Inman's solution (an elliptic cycloid) against the three arbitrary curves proposed here (A, B and C). Solely the shoaling part of the beach profile (or shorerise profile) is considered. The measured profile is that labelled `a) Survey Range PN 1180 March 1981' in fig. 8 of Jenkins and Inman \cite{Jenkins}}
	\label{fig:profile}
\end{figure}

\begin{table}
	\begin{center}
    \begin{tabular}{lcrrrrr}
     & \multicolumn{6}{c}{\textbf{Jenkins \& Inman (2006) profiles}} \\
    \multicolumn{1}{c}{\textbf{}} & \textbf{a} & \multicolumn{1}{c}{\textbf{b}} & \multicolumn{1}{c}{\textbf{c}} & \multicolumn{1}{c}{\textbf{d}} & \multicolumn{1}{c}{\textbf{e}} & \multicolumn{1}{c}{\textbf{f}} \\
    \textbf{value of $e$} & \multicolumn{1}{r}{0.70} & 0.66 & 0.74 & 0.77 & 0.70 & 0.76 \\
    \textbf{corresponding $n$} & \multicolumn{1}{r}{0.96} & 0.67 & 1.31 & 1.56 & 0.95 & 1.48
    \end{tabular}
	\caption{Value of the calibration parameter $e$ reported by Jenkins and Inman \cite{Jenkins} for the shoaling part of each of the six profiles considered (see fig. 8 in their paper), and corresponding $n$ according to eq. \eqref{eq:e-n}.}
	\label{table_n}
	\end{center}
\end{table}

\begin{table}
	\begin{center}
	\begin{tabular}{lrrrrrr}
		& \multicolumn{6}{c}{\textbf{Jenkins \& Inman (2006) profiles}} \\
		\multicolumn{1}{c}{\textbf{}} & \multicolumn{1}{c}{\textbf{a}} & \multicolumn{1}{c}{\textbf{b}} & \multicolumn{1}{c}{\textbf{c}} & \multicolumn{1}{c}{\textbf{d}} & \multicolumn{1}{c}{\textbf{e}} & \multicolumn{1}{c}{\textbf{f}} \\
		\textbf{Curve A} & 1.30 & 1.21 & 1.29 & 1.38 & 1.23 & 1.38 \\
		\textbf{Curve B} & 1.06 & 1.08 & 1.01 & 0.83 & 1.07 & 0.94 \\
		\textbf{Curve C} & \multicolumn{1}{l}{0.97} & \multicolumn{1}{l}{1.00} & \multicolumn{1}{l}{0.94} & \multicolumn{1}{l}{0.78} & \multicolumn{1}{l}{1.00} & \multicolumn{1}{l}{0.88}
	\end{tabular}
	\caption{Ratio of the functional $J[x(h)]$ (eq. \ref{eq:functional}) obtained by the curves proposed here to that yielded by Jenkins and Inman's solution; i.e. ratio of $J[x(h)=$ curve shown in left column$]$ to $J[x(h)=$ Jenkins and Inman's solution$]$. Shorerise profiles are those shown in fig. 8 of \cite{Jenkins}.}
	\label{table}
	\end{center}
\end{table}

For comparison, we use the six profiles (a, b, ... , f) shown in fig. 8 of Jenkins and Inman \cite{Jenkins}. 
Table \ref{table_n} shows the values of $e$ reported for the shorerise profile by \cite{Jenkins}, and corresponding $n$ according to eq. \eqref{eq:e-n}. 
These are the values of $n$ that we use in \eqref{eq:functional} and in Curve B for comparison against Jenkins and Inman's solution.

Table \ref{table}, which gives the ratio of $J[x(h)]$ (eq. \ref{eq:functional}) yielded by Curves A, B and C to that obtained from Jenkins and Inman's solution, illustrates the following points:

\begin{itemize}
	\item The solution by Jenkins and Inman does not maximise the functional (\ref{eq:functional}). Other curves yield greater values of $J[x(h)]$; most notably, the linear profile (Curve A). (see also section 5 in \cite{Maldonado})
	\item The solution by Jenkins and Inman does not minimise the functional (\ref{eq:functional}) either, and so we can state, more generally, that it does not represent an extremum (see values $<1$ for Curves B and C). 
\end{itemize}

The claim by Jenkins and Inman \cite{Jenkins} that an equilibrium beach profile described by an elliptic cycloid maximises the rate of energy dissipation of non-breaking waves, in turn related to the integral (\ref{eq:integral}), is therefore incorrect. 

\section{Conclusions}

Jenkins and Inman \cite{Jenkins} claimed that an equilibrium beach profile described by an elliptic cycloid \textit{maximises} the rate of energy dissipation of both breaking and non-breaking waves. 
However, focusing on non-breaking waves only, we have shown here that i) other curves (e.g. a line) yield larger rates of energy dissipation as formulated by Jenkins and Inman themselves, thus invalidating their claim; and ii) the solution proposed by Jenkins and Inman to the associated variational problem invites further scrutiny given the missing information which is crucial to recover it.

\appendix
\section{Appendix}

\hspace{0.5cm} Incomplete elliptic integrals of the first kind, $F(\varphi , k)$, and of the second kind, $E(\varphi , k)$:
\begin{equation}
F(\varphi , k) = \int_{0}^{\varphi} \frac{d \theta}{\sqrt{1 - k^2 \sin^2 \theta}} \; \; \; \; \textup{and} \; \; \; \; E(\varphi , k) = \int_{0}^{\varphi} \sqrt{1 - k^2 \sin^2 \theta} d \theta .
\end{equation} 

Complete elliptic integrals of the first kind, $K(k)$, and of the second kind, $E(k)$:
\begin{equation}
K(k) = \int_{0}^{\frac{\pi}{2}} \frac{d \theta}{\sqrt{1 - k^2 \sin^2 \theta}} \; \; \; \; \textup{and} \; \; \; \; E(k) = \int_{0}^{\frac{\pi}{2}} \sqrt{1 - k^2 \sin^2 \theta} d \theta .
\end{equation} 

\textbf{Acknowledgements:} MU wishes to acknowledge the support received via the programme Erasmus+, which allowed her to complete an internship at the University of Southampton, where the core of this work was developed. Codes employed in this paper can be found at \url{https://github.com/sergio-maldonado/on-JI2006-solution}.


\end{document}